\documentclass[reqno]{amsart}

\usepackage[usenames,dvipsnames]{color}
\usepackage[bookmarks,colorlinks,breaklinks]{hyperref}  
\usepackage{doi,url}

\definecolor{dullmagenta}{rgb}{0.4,0,0.4}   
\definecolor{darkblue}{rgb}{0,0,0.4}
\hypersetup{linkcolor=dullmagenta,citecolor=blue,filecolor=dullmagenta,urlcolor=darkblue} 

\usepackage{amssymb,yhmath}
\usepackage{enumerate}

\usepackage{color}

\newcommand{\eq}[1]{\eqref{#1}}

\newcommand{\bu}{\textcolor{blue}}

\newtheorem{theorem}{Theorem}[section]

\newtheorem{lemma}[theorem]{Lemma}

\newtheorem{remark}[theorem]{Remark}

\newtheorem*{remark*}{Remark}

\numberwithin{equation}{section}

\newenvironment{acknowledgement}{\emph{Acknowledgement.}}

\DeclareMathOperator{\supp}{supp}

\DeclareMathOperator{\sech}{sech}

\def\Re{\mathop\mathrm{Re}}

\providecommand{\wtilde}[1]{\widetilde{#1}}
\providecommand{\what}[1]{\widehat{#1}}

\newcommand{\pr}{\prime}
\newcommand{\Xicl}{\Xi^{\mathrm{cl}}}
\def\tuv{\mathop\mathcal{T}}
\def\wlim{\mathop\textup{w$^*$-lim}}
\def\wwlim{\mathop\textup{w-lim}}
\def\slim{\mathop\textup{s-lim}}
\newcommand\bdelta{\boldsymbol{\delta}}

\newcommand\R{\mathbb R}

\newcommand\E{\mathbb E}

\newcommand{\CC}{\mathbb{C}}
\newcommand{\EE}{\mathbb{E}}

\newcommand{\RR}{\mathbb{R}}

\newcommand{\ZZ}{\mathbb{Z}}

\newcommand{\cE}{\mathcal{E}}

\newcommand{\cL}{\mathcal{L}}

\newcommand{\cF}{\mathcal{F}}
\newcommand{\cK}{\mathcal{K}}

\newcommand{\cT}{\mathcal{T}}

\newcommand{\cH}{\mathcal{H}}

\newcommand\e{\mathrm{e}}
\renewcommand{\d}{\mathrm{d}}

\newcommand{\dX}{\dot{X}_{1}}   
  
\def\llangle{\langle\mkern-4mu\langle}
\def\rrangle{\rangle\mkern-4mu\rangle} 

\newcommand\Chi{\raisebox{.2ex}{$\chi$}}

\newcommand{\abs}[1]{\left\lvert #1 \right\rvert}
\newcommand{\norm}[1]{\left\lVert #1 \right\rVert}

\newcommand{\set}[1]{\left\{ #1 \right\}}
\newcommand{\pa}[1]{\left( #1 \right)}

\newcommand{\tnorm}[2][]{\ifx @#1@\def\myleft{\left}%
          \def\myright{\right}%
     \else\def\myleft{\csname#1l\endcsname}%
          \def\myright{\csname#1r\endcsname}%
          \ifx\myleft\normall\def\myleft{\relax}\def\myright\relax\fi%
     \fi%
     \myleft|\!\myleft|\!\myleft| #2 \myright|\!\myright|\!\myright|}

\newcommand\beq{\begin{equation}}
\newcommand\eeq{\end{equation}}

\newcommand{\qtx}[1]{\quad\text{#1}\quad}
\newcommand{\sqtx}[1]{\;\text{#1}\;}
\newcommand{\mqtx}[1]{\;\;\text{#1}\;\;}

\begin{document}

\title
[Ac-conductivity and  energy absorption for the Anderson model]
{Ac-conductivity and  electromagnetic energy absorption for the Anderson model in linear response theory}

\author{Abel Klein}
\address[A. Klein]{University of California, Irvine;
Department of Mathematics;
Irvine, CA 92697-3875,  USA}
 \email{aklein@uci.edu}

\author[Peter \ M\"uller]{Peter M\"uller}
\address[P.\ M\"uller]{Mathematisches Institut,
  Ludwig-Maximilians-Universit\"at,
  Theresienstra\ss{e} 39,
  80333 M\"unchen, Germany}
\email{mueller@lmu.de}

\thanks{A.K. was  supported in part by the NSF under grant DMS-1301641.}

\dedicatory{Dedicated to Leonid A.\ Pastur on the occasion of his 75th
   birthday}

\begin{abstract} 
We continue our study of the ac-conductivity in linear response theory for the Anderson  model   using the conductivity measure.  We establish further properties of the conductivity measure, including nontriviality at nonzero temperature, the high temperature limit, and asymptotics with respect to the disorder. We also calculate the electromagnetic energy absorption in linear response theory in terms of the conductivity measure.
 \end{abstract}

\maketitle

\tableofcontents

\section{Introduction}

We continue our study of the ac-conductivity in linear response theory for the Anderson  model following  \cite{KLM,KMu}, where we  introduced the concept of a conductivity measure. The conductivity
measure $ \Sigma_{\mu}^{T}(\d \nu)$ at  absolute temperature $T\ge 0$ and   Fermi level  (chemical
potential) $\mu\in\RR$    is a finite positive even Borel measure on frequency space  ($\nu$ denotes  the frequency of the applied electric field). If $ \Sigma_{\mu}^{T}(\d
\nu)$ was known to be an absolutely continuous measure, the in-phase
(or active) conductivity $ \Re \sigma_{\mu}^T(\nu)$ would then be
well-defined as its density. The conductivity measure $
\Sigma_{\mu}^{T}(\d \nu)$ is  an analogous concept to the density
of states measure, whose formal density is the
density of states. The Mott formula proved in  \cite{KLM} is a statement about the asymptotic behavior of $\Sigma_{\mu}^{0}([0,\nu])$ as $\nu\downarrow 0$ for a Fermi level $\mu$ within the region of complete localization.

In this article we establish further properties of the conductivity measure, including nontriviality at nonzero temperature,   the high temperature limit, and asymptotics with respect to the disorder. We also calculate the electromagnetic energy absorption in linear response theory in terms of the conductivity measure.

 This work is motivated by
a series of papers by 
Bru, de Siqueira Pedra and Kurig\footnote{We thank Jean-Bernard Bru, Walter de Siqueira Pedra and Carolin Kurig for communicating their work to us at an early stage and for stimulating discussions.}  \cite{BPK1,BPK2,BPK3,BPK4}, who  study the linear response  of an infinite system of free fermions in the lattice  to an electric field given by    a time and space dependent   potential.  They assume the presence of impurities, and take the one-particle Hamiltonian to  be  the Anderson model.  They show that at nonzero temperature ($T>0$)   the electromagnetic energy absorbed (`heat production')  in linear response theory is given in terms of an ac-conductivity measure. They also   derive asymptotics for this  ac-conductivity measure with respect to the disorder.  Although the  BPK (for Bru, de Siqueira Pedra and Kurig) mathematical setting is very different from ours (see, e.g.,  the discussion in \cite[Section~2.3]{BPK3}), there are clear   similarities  
 between their work and ours.

The Anderson model is described by the random
Schr\"odinger operator $H$, a measurable map $\omega \mapsto
H_{\omega}$ from a probability space $(\Omega,\mathbb{P})$ (with
expectation $\mathbb{E}$) to bounded self-adjoint operators on
$\ell^2(\ZZ^d)$, given by
\begin{equation}
   \label{AND}
   H_\omega := - \Delta + V_\omega ,
\end{equation}
where  $\Delta$ is the centered discrete  Laplacian
\begin{equation}
  (\Delta \varphi)(x):= -  \sum_{{y\in\ZZ^d; \, |x-y|=1}} \varphi(y)  
\qquad
  \text{for} \quad   \varphi\in\ell^2(\ZZ^d), \quad x \in \ZZ^{d},
\end{equation}
and the random potential $V_\omega$ consists of independent, identically
distributed random variables $\{V_\omega(x) ; x \in \ZZ^d\}$ on
$(\Omega,\mathbb{P})$, such that the common single site probability
distribution $\xi$ is nondegenerate with  compact support, say
\beq\label{vpm}
\set{v_-,v_+}\in \supp \xi \subset [v_-,v_+], \qtx{where} -\infty < v_- < v_+ < \infty,
\eeq
 and has a bounded density $\rho \in \mathrm{L}^{\infty}(\RR)$.
 
The {Anderson Hamiltonian} $H$ given by \eqref{AND} is
$\mathbb{Z}^d$-ergodic.  It follows that its spectrum is nonrandom:  there exists a set $\mathfrak{S}\subset \R$ such that
$\sigma(H_\omega)= \mathfrak{S}$ with probability one \cite{Pa74,Pa80}.  Moreover, the pure point,
absolutely continuous, and singular continuous components of $\sigma
(H_\omega)$  are also nonrandom, i.e., equal to fixed sets $\mathfrak{S}_{\mathrm{pp}}, \mathfrak{S}_{\mathrm{ac}}, \mathfrak{S}_{\mathrm{sc}}$  with probability one,
and $\mathfrak{S}= [-2d,2d] + \supp \xi$ (see \cite{KiM, CL, PF,Ki}).  In particular, setting
$E_-:= -2d + v_-$ and $E_+ := 2d + v_+$, we have
\beq\label{Epm}
-\infty < E_{-}=\inf  \mathfrak{S}< E_+= \sup \mathfrak{S}<\infty, \mqtx{so} \set{E_-,E_+} \subset \mathfrak{S}\subset [E_-,E_+].
\eeq

We start by reviewing the derivation of electrical ac-conductivities within linear response theory for the Anderson model following \cite{BGKS,KLM,KMu}.  At  time $t=-\infty$, the system is  in
thermal equilibrium at absolute temperature $T\ge 0$ and chemical
potential $\mu\in\RR$. In the single-particle Hilbert space $\ell^{2}(\ZZ^{d})$ this equilibrium
state is given by the random operator $f_{\mu}^{T}(H)$, where
\begin{equation}
  \label{fermi}
  f_{\mu}^{T}(E) := 
  \begin{cases}
    \pa{ \e^{ \frac{E-\mu}T } +1}^{-1} & \text{if} \quad  T>0  \smallskip \\
    \Chi_{]-\infty, \mu]}(E) &\text{if} \quad  T=0
  \end{cases}
  \end{equation}
is the Fermi function. By $\Chi_{B}$ we denote the characteristic 
function of the  set $B$. A spatially homogeneous, time-dependent
electric field $\mathbf{E}(t)$ is then introduced adiabatically:
Starting at time $t= -\infty$, we switch on the (adiabatic) electric
field $\mathbf{E}_{\eta}(t):= \e^{\eta t}\mathbf{E}(t)$ with $\eta
>0$, and then let $\eta \to 0$. 

In view of isotropy we assume
without loss of generality that the electric field points in the
$x_{1}$-direction: $\mathbf{E}(t)=\mathcal{E}(t) \widehat{x}_{1} $,
where $\mathcal{E}(t)$ is the (real-valued) amplitude of the electric
field, and $\widehat{x}_{1}$ is the unit vector in the
$x_{1}$-direction.  
We assume 
\beq\label{assumption1}
\cE(t) =  \int_{\RR}\!\d \nu \; \e^{i\nu t}\widehat{\cE}(\nu), \mqtx{where}  \widehat{\cE} \in C(\RR)\cap \mathrm{L}^1(\RR)  \mqtx{with}
   \widehat{\cE}(\nu)=\overline{ \widehat{\cE}(-\nu)}.
   \eeq
 Note that \eq{assumption1} implies that  ${\cE} \in C(\RR)\cap \mathrm{L}^\infty(\RR)$  and is real-valued.

For each $\eta>0$  this procedure results in a time-dependent random Hamiltonian
\begin{equation}\label{Homegaeta}
   H_{\omega}({\eta, t}) := G(\eta,t) H_{\omega} G(\eta,t)^{*}, \quad \text{with} \quad
   G(\eta,t) := \e^{iX_{1} \int_{-\infty}^{t}\d s\,
     \e^{\eta s} \cE(s)},
\end{equation}
where $X_{1}$ stands for the operator of multiplication by the first
coordinate of the electron's position.  (The time-dependent, bounded Hamiltonian $ H_{\omega}({\eta, t})$ is  gauge equivalent to $H_{\omega} + \e^{\eta t} \cE(t)
X_{1}$; this choice of gauge is discussed in \cite[Section~2.2]{BGKS}.)  The state of the system is described  at time $t$ by the
random operator $\varrho_{\mu,\omega}^T(\eta,t)$, the solution to
the Liouville equation
\begin{equation}\label{Liouvilleeq}
   \left\{
     \begin{array}{l}
     \mbox{$i$} \partial_t \varrho_{\mu,\omega}^T(\eta,t) =
     [H_{\omega}(\eta,t),\varrho_{\mu,\omega}^T(\eta,t)]
       \\
       \displaystyle\lim_{t \to  -\infty}\varrho_{\mu,\omega}^T(\eta,t)=
       f_{{\mu}}^{T}(H_{\omega})
   \end{array}
   \right.  .
\end{equation}

The adiabatic electric field generates a time-dependent electric
current,  also oriented along
the first coordinate axis. In the Schr\"odinger picture  it has  amplitude 
\begin{equation}
   \label{curdef}
   J_{\eta}(t;\mu,T,\mathcal{E}) = -  \cT \bigl( \varrho_{\mu,\omega}^{T} 
(\eta,t)
   \dot{X}_{1}(\eta,t)\bigr),
\end{equation}
where $\cT$ is the trace per unit volume, defined below Eq.\ \eqref{scalarproduct}, and $ \dot{X}_{1}(\eta,t)$ is the first
component of the time-dependent velocity operator:
\begin{equation}\label{dotX}
 \dot{X}_{1}(\eta,t) :=
G(\eta,t)\dX G(\eta,t)^{*}, \qtx{where}  \dX := i [H_{\omega}, X_{1}] = i [-\Delta, X_{1}].
\end{equation}
The \emph{adiabatic linear-response current} is defined as
\begin{equation}
   \label{lin-cur}
   J_{\eta,\mathrm{lin}}(t;\mu,T,\cE) := \frac{\d}{\d\alpha}\,
   J_{\eta }(t;\mu,T,\alpha\cE)\big|_{\alpha=0}.
\end{equation}

The detailed analysis in \cite{BGKS}  gives a
mathematical meaning to the formal procedure leading to
\eqref{lin-cur}, for fixed temperature $T\ge 0$ and chemical potential
$\mu \in \RR$, when  the corresponding thermal equilibrium random
operator $ f_{\mu}^{T}(H_\omega)$ satisfies the condition
\begin{equation} 
  \label{assumpIntro} 
  \E\big\{ \big\| X_{1}\,
      f_{\mu}^{T}(H_{\omega}) \delta_0\big\|^2 \big\} < \infty ,
\end{equation}
where $\set{\delta_a}_{a \in \ZZ^d}$ is the canonical orthonormal basis in $\ell^2(\ZZ^d)$: $\delta_a(x):=1 $ if $x=a$ and $\delta_a(x):=0$ otherwise.  (This condition appears in \cite{BES}.)   This analysis requires the mathematical framework of normed spaces of measurable covariant
operators  given in \cite[Section~3]{BGKS}, and described in   \cite[Section~3]{KLM} and \cite[Appendix~A]{KMu}.
 Here we will only give a short (and informal) review.  By $\cK_{2}$ we denote
the Hilbert space of measurable covariant operators $A$ on
$\ell^2(\ZZ^d)$, i.e., measurable, covariant maps $\omega \mapsto
A_{\omega}$ from the probability space $(\Omega,\mathbb{P})$ to
operators on $\ell^2(\ZZ^d)$, with inner product
\begin{equation}
  \label{scalarproduct}
  \llangle A, B\rrangle :=  \ \EE \bigl\{\langle
  A_{\omega}\delta_{0}, B_{\omega}  
  \delta_{0}\rangle
  \bigr\} =  \cT \set{A^{*} B}
\end{equation}
and norm $ \tnorm{{A}}_2:= \sqrt{ \llangle A, A\rrangle}$. Here $\cT$,
given by $\cT(A) := \EE \{\langle\delta_{0}, A_{\omega}
\delta_{0}\rangle \}$, is the trace per unit volume.  The Liouvillian
$\cL$ is the (bounded in the case of the Anderson model) self-adjoint
operator on $\cK_2$ given by the commutator with $H$: 
\beq \pa{\cL
  A}_\omega:= [H_\omega, A_\omega].  
\eeq 
We also introduce the operators $\cH_L$ and $ \cH_R$ on $\cK_2$ that are given by left and right multiplication
by $H$: 
\beq \pa{\cH_L A}_\omega:= H_\omega A_\omega \qtx{and}
\pa{\cH_R A}_\omega:= A_\omega H_\omega.  
\eeq 
They are commuting, bounded (for the Anderson Hamiltonian),
self-adjoint operators on $\cK_2$, anti-unitarily equivalent, and $\cL =\cH_L - \cH_R$. 
It is easy to see that 
\beq\label{spectrumHLR}
\sigma(\cH_L)=\sigma(\cH_R) \subset \mathfrak{S},\qtx{so} 
\sigma (\cL) = \sigma(\cH_L) -\sigma(\cH_R)\subset \mathfrak{S}- \mathfrak{S}.
\eeq
 It follows from the Wegner
estimate for the Anderson model that  the operators
$\cH_L$ and $ \cH_R$ have purely absolutely continuous spectrum \cite[Lemma~1]{KMu}.  For each $T \ge
0$ and $\mu \in \RR$ we consider the bounded self-adjoint operator
$\cF_\mu^T$ in $\cK_2$ given by
 \begin{equation} \label{defFmuT}
  \cF_\mu^T := f_{\mu}^{T}(\cH_{L}) - f_{\mu}^{T}(\cH_{R})   , \quad \text{i.e.}, \quad \pa{\cF_\mu^T A}_\omega= [f_\mu^T(H_\omega),A_\omega]. 
  \end{equation}

In this formalism the  condition \eq{assumpIntro} can  be rewritten as
\begin{equation}\label{defY}
Y_{\mu}^{T} := i [X_{1}, f_{\mu}^{T}(H)] \in \cK_{2},
\end{equation}
 which is always true for $T>0$ and arbitrary $\mu \in \RR$,
since in this case $f_{\mu}^{T}(H)= g(H)$ for some Schwartz function $g \in \mathcal{S}(\RR^d)$
(\cite[Remark~5.2(iii)]{BGKS}).  We set
\begin{align}
  \Xi_0 := 
  \set{\mu \in \RR; \quad Y_{\mu}^{0}  \in \cK_{2}}.
\end{align}
For the same reason as when $T>0$, we have  $\mu
\in \Xi_0 $ if either $\mu \notin \mathfrak{S}$ or $\mu$ is the left edge of a
spectral gap for $H$.  Moreover, letting $\Xicl$ denote the region of complete localization (see \cite{GKjsp}),  defined as the region of
validity of the multiscale analysis, or equivalently, of the fractional moment
method, we have  (see  \cite{AG,GKjsp})
\beq \label{XiclXi0}
\Xicl \subset \Xi_0 .
\eeq
 We refer to  \cite[Appendix~B]{KMu} for a precise definition. Note $\R\setminus  \mathfrak{S}\subset \Xicl$ and that  $\Xicl$ is an open set by its definition.  Note also that
   for $\mu \in \Xicl$ the Fermi projection $f_{\mu}^{0}(H)$ satisfies a
  much stronger condition than \eq{assumpIntro}, namely exponential decay of
  its kernel \cite[Theorem~2]{AG}.  Conversely, fast enough polynomial decay
  of the kernel of the Fermi projection for all energies in an interval
  implies complete localization in the interval \cite[Theorem~3]{GKjsp}.

If $ Y_{{{\mu}}}^{T} \in \cK_{2}$, 
an inspection of the proof of   \cite[Thm.~5.9]{BGKS} shows that  the adiabatic linear-response current \eqref{lin-cur} is  well
defined for all $t\in\RR$, and given by (see \cite[Eq.~(2.18)]{KMu})
\begin{align}
     \label{lin-cur2}
     J_{\eta,\mathrm{lin}}(t;{{\mu}},T,\cE) &= \tuv \left\{
       \int_{-\infty}^{t}\!\d s\; \e^{\eta s} \cE(s) \dX
       \e^{- i (t-s) \cL} Y_{{{\mu}}}^{T}
            \right\}\\ \notag
         &=    \int_{-\infty}^{t}\!\d s\; \e^{\eta s} \cE(s) \llangle \dot{X}_{1},
      \e^{- i (t-s) \cL}  Y_{{{\mu}}}^{T} \rrangle   .
   \end{align}
   
 We introduced the conductivity measure in \cite{KLM,KMu} to rewrite  \eq{lin-cur2}.  If $ Y_{{{\mu}}}^{T} \in \cK_{2}$, 
  the  {(ac-)conductivity measure} ($x_{1}$-$x_{1}$ component) at temperature $T$ and Fermi level ${{\mu}}$
  is defined by
  \begin{equation}
     \label{kubo-expr}
     \Sigma_{{{\mu}}}^{T}(B) := \pi \llangle \dot{X}_{1},
     \Chi_{B}(\mathcal{L}) Y_{{{\mu}}}^{T} \rrangle  \quad \text{for all Borel sets $B\subset\mathbb{R}$}.
\end{equation}
We proved that $\Sigma_{{{\mu}}}^{T}$  is a  finite positive even Borel measure on the real line \cite[Theorem~1]{KMu}. Thus  \eq{lin-cur2} can be rewritten as 
\begin{align}\label{lin-cur22}
 J_{\eta,\mathrm{lin}}(t;{{\mu}},T,\cE)&=   \frac{1}{\pi} \int_{-\infty}^{t}\!\d s\; \e^{\eta s} \cE(s)  \int_{\mathbb{R}}\!\Sigma_{{\mu}}^{T}(\d\lambda)\  \e^{- i (t-s) \lambda}\\
& =  \label{lin-cur25}
  \e^{\eta t}  \int_{\RR}
  \!\d\nu \,\e^{i\nu t} 
  \sigma_{{\mu}}^{T}(\eta,\nu) \,
  \widehat{\cE}(\nu),
\end{align}
where 
\begin{equation}
  \sigma_{{\mu}}^{T}(\eta,\nu) := - \frac{i}{\pi}
  \int_{\mathbb{R}}\!
  \Sigma_{{\mu}}^{T}(\d\lambda)\; \frac{1}{\lambda +\nu - i\eta}.
\end{equation}
We then defined
 the adiabatic in-phase linear-response current  by
\begin{equation}
  J_{\eta,\mathrm{lin}}^{\mathrm{in}}(t;{{\mu}},T,\cE) := \e^{\eta t}
  \int_{\RR} \!\d\nu \,\e^{i\nu t}
  \bigl( \Re\sigma_{{\mu}}^{T}(\eta,\nu)\bigr) \,
  \widehat{\cE}(\nu).
\end{equation}
Turning off the adiabatic switching, we obtained a simple expression for
the in-phase linear-response current in terms of the conductivity
measure, given by
\begin{equation}\label{Jin}
  J_{\mathrm{lin}}^{\mathrm{in}}(t; 
  {\mu},T,\cE):=   \lim_{\eta\downarrow0} J_{\eta,\mathrm{lin}}^{\mathrm{in}}(t; 
 {\mu},T,\cE) =
  \int_{\mathbb{R}}\!\Sigma_{{\mu}}^{T}(\d\nu)\; \e^{i\nu t}
  \widehat{\mathcal{E}}(\nu).
  \end{equation}
This  derivation of the in-phase linear-response current
 is valid as long as either $T>0$ or ${\mu} \in
\Xi_0 $, so we can guarantee \eq{defY}. In addition, we proved \cite[Eq.\ (2.31)]{KMu} that
\beq \label{Jin0}
 J_{\mathrm{lin}}^{\mathrm{in}}(t; 
{\mu},0,\cE)=   \lim_{T \downarrow 0}  J_{\mathrm{lin}}^{\mathrm{in}}(t; 
 {\mu},T,\cE) \quad \text{for all} \;\; {\mu} \in \Xi_0.
\eeq

In  \cite{KMu} we also extended the definition of the conductivity measure at $T=0$ to arbitrary Fermi level  ${\mu}$.  Given $T>0$ and ${\mu} \in \R$, 
we decompose $\Sigma_{{\mu}}^{T}$ as
\beq
\label{decompose}
\Sigma_{{\mu}}^{T}= \Sigma_{{\mu}}^{T}\pa{\set{0}}\bdelta_{0} + \pa{\Sigma_{{\mu}}^{T} -   \Sigma_{{\mu}}^{T}\pa{\set{0}}\bdelta_{0}}= \Psi\pa{
        \pa{-f_{{\mu}}^{T}}^{\pr}}\bdelta_{0} + \Gamma_{{\mu}}^{T},
\eeq
where  $\bdelta_{0}$ is the Dirac measure at $0$, and  $\Psi$ and $\Gamma_{{\mu}}^{T}$ are  finite  positive Borel measures on $\R$ given by 
\begin{align}
  \label{defPsi}
      \Psi(B)& :=\pi  \llangle \dX,\chi_{\{0\}}(\cL)
      \,\Chi_{B}(\cH_L)\dX\rrangle, \\
      \label{defGamma1}
      \Gamma_{{\mu}}^{T}(B) &:= \pi  \llangle \pa{ -
        \cL_\perp^{-1}\cF_{{\mu}}^T}^{\frac 1 2} \dX, \Chi_{B}(\cL)\pa{
        - \cL_\perp^{-1}\cF_{{\mu}}^T}^{\frac 1 2} \dX\rrangle. 
\end{align}
Here $\cL_\perp^{-1} $ denotes the pseudo-inverse of $\cL$ (i.e.,  
$
\cL_\perp^{-1} := g(\cL)$ where
$g(t) := \frac 1 t$ if $t\not= 0$ and $g(0):=0$)
 and $\cF_{{\mu}}^T$ is given in \eq{defFmuT}. Note that $\Gamma_{{\mu}}^{T}(\{0\})=0$.  
(In \eqref{decompose} we used the short-hand notation $\Phi(h) := \int_{\RR}\Phi(\d\lambda) \,h(\lambda)$ for the integral of a function $h$ with respect to the measure $\Phi$.)

 We let $\mathcal{M}(\RR)$ denote the vector
space of complex Borel measures on $\RR$, with $\mathcal{M}_+(\RR)$
being the cone of finite positive Borel measures, and with
$\mathcal{M}_+^{(\e)}(\RR)$ the finite positive even Borel measures.
We recall that $\mathcal{M}(\RR)=C_0(\RR)^*$, where $C_0(\RR)$ denotes
the Banach space of complex-valued continuous functions on $\RR$
vanishing at infinity with the sup norm.  We will use three locally
convex topologies on $\mathcal{M}(\RR)$. The first is the weak$^*$
topology (also called the vague topology), induced by the linear functionals $\set{\Gamma \in
  \mathcal{M}(\RR) \mapsto \Gamma(g); \; g \in C_0(\RR)}$.  The second is the weak topology,
  defined in the same way as the weak$^*$
topology but with $C_{\mathrm{b}}(\RR)$, the bounded continuous functions on $\R$, substituted for  $C_0(\RR)$.
The third is the strong topology,  induced by the linear functionals $\set{\Gamma \in
  \mathcal{M}(\RR) \mapsto \Gamma(B); \; B\subset \R \, \;\text{Borel set}}$.
    We will write
$\wlim$, $\wwlim$, and $\slim$, to denote limits in the  weak$^*$, weak, and strong topology, respectively.

 We proved \cite[Theorem~2]{KMu} that the measure $\Psi$ from \eqref{defPsi} is absolutely continuous with  density $\psi$, where $\psi(E)=0$ on  $ \Xi_{0}$ since  $\supp \Psi\subset \overline{ \RR \setminus \Xi_{0}}\subset  \RR \setminus \Xicl$.  We defined
\begin{equation}
     \label{kubo-expr2}
     \Sigma_{{\mu}}^{0} :=\psi({\mu})\bdelta_{0} +  \Gamma_{{\mu}}^{0},
\end{equation}
which coincides with the previous definition for ${\mu} \in \Xi_0$, and  yields 
\beq\label{weakT0}
 \Sigma_{{\mu}}^{0}(\d\nu)= \wlim_{T\downarrow 0}
\Sigma_{{\mu}}^{T}(\d\nu)\quad \text{for Lebesgue-a.e.\ ${\mu}\in\RR$ }.
\eeq
Moreover, for all for all temperatures  $T>0$ and and Fermi levels
${\mu}\in \R$ we have 
 \beq
     \label{convolution}
        \Sigma_{{\mu}}^{T} = \pa{ \pa{-f_{0}^{T}}^\prime
          \ast \Sigma_{\bullet}^0}({\mu}) , \quad\text{i.e.,}\quad
        \Sigma_{{\mu}}^{T}(B)  = \int_{\RR}\!\d E \; (-f_{{\mu}}^{T})'(E)
        \; \Sigma_{E}^{0}(B) ,
    \eeq 
which justifies the extension   of  the definition of  in-phase linear-response current by
\beq 
J_{\mathrm{lin}}^{\mathrm{in}}(t; {\mu},0,\cE):= \int_{\mathbb{R}}\!\Sigma_{{\mu}}^{0}(\d\nu)\; \e^{i\nu t}
  \widehat{\mathcal{E}}(\nu)= 
\lim_{T\downarrow 0} J_{\mathrm{lin}}^{\mathrm{in}}(t; {\mu},T,\cE) 
\sqtx{for Lebesgue-a.e.} {\mu}\in\RR .  
\eeq
(We refer to \cite{KMu} for full details and proofs.)

In Section~\ref{sechighT} we establish nontriviality of the conductivity measure at $T>0$ and prove that $\Sigma_{{\mu}}^{T} \to 0$  strongly as $T\to \infty$. In Section~\ref{secdisorder} we introduce a disorder parameter $\lambda$ and study the weak$^*$ limit of $\Sigma_{{\mu},\lambda}^{T}$
as $\lambda \to 0$  and the strong limit of $\Sigma_{{\mu},\lambda}^{T}$ as $\lambda \to \infty$.  Finally, in  Section~\ref{secwork} we derive an expression, in linear response theory,    for the 
total electromagnetic energy $W^T_{\mathrm{{\mu},lin}}(\cE)$ absorbed by the solid  during all times in terms of the conductivity measure  $\Sigma_{{\mu}}^{T}$:  
\beq
W^T_{\mathrm{{\mu},lin}}(\cE)
		= 2\pi\int_{\RR}\!\Sigma_{{\mu}}^{T}(\d\nu)  \, |\what{\cE}(\nu)|^{2} \ge 0.
		\eeq

\bigskip
\begin{acknowledgement} 
  This paper is dedicated to {Leonid A.\ Pastur} on the occasion of his 75th birthday.
  Pastur is a founding father of the theory of random Schr\"odinger
  operators; of particular relevance to this paper is his work on the
  electrical conductivity, e.g.,
  \cite{BePa70,Pas71,Pas73,LiGr88,KP,Pas99a,Pas99b,KiLe03}. 
   \end{acknowledgement}

\section{Nontriviality and high temperature limit  of the conductivity measure}\label{sechighT}

We  prove nontriviality of the conductivity measure for $T>0$  and strong convergence to $0$ as $T \to \infty$.
In the BPK setting,  nontriviality of the ac-conductivity measure   for sufficiently large  $T$ is shown in  \cite[Theorem~4.7]{BPK4}.

\begin{theorem}\label{thmTinfty} Let $T>0$ and  ${\mu}\in\RR$. Then  the conductivity  measure $\Sigma_{{\mu},\lambda}^{T}$ and the measure  $\Gamma_{{\mu},\lambda}^{T}$ are nontrivial:  
\beq
\Sigma_{{\mu}}^{T}(\RR)\ge  \Gamma_{{\mu}}^{T}(\RR) >0.
\eeq
  Moreover,
we have 
\beq
\slim_{T \to \infty} \Sigma_{{\mu}}^{T} =\slim_{T \to \infty} \Gamma_{{\mu}}^{T} =0 \qtx{for all} {\mu}\in\RR. 
\eeq
\end{theorem}

\begin{remark}\label{SigmasigmacL} It follows from \eq{kubo-expr} and \eq{defGamma1} that  
\beq
  \Sigma_{{\mu}}^{T}\big(\RR \setminus \sigma (\cL)\big)=0 \qtx{and} 
  \Gamma_{{\mu}}^{T}\Big(\big(\RR \setminus \sigma (\cL)\big)\cup \{0\}\Big)=0.
  \eeq 
Thus only  frequencies $\nu\in \sigma (\cL)\subset \mathfrak{S}- \mathfrak{S}$ (recall \eq{spectrumHLR}) contribute to $ \Sigma_{{\mu}}^{T}$ and $\Gamma_{{\mu}}^{T}$.  It follows from Theorem~\ref{thmTinfty} that
\beq
\Sigma_{{\mu}}^{T}\big( \sigma (\cL)\big)\ge \Gamma_{{\mu}}^{T}\big( \sigma (\cL)\setminus \set{0}\big)>0 \qtx{for all} T>0 \qtx{and} \mu \in \R.
\eeq 
\end{remark}

To prove the theorem we introduce  the 
 finite Borel measure $\Upsilon$ on $\RR$ given by
\beq\label{defUpsilon}
\Upsilon(B) := \llangle \dX,\Chi_{B\setminus \{0\}}(\cL)   \dX\rrangle  \qtx{for all Borels sets} B \subset \R.
 \eeq
Since 
  $[\dX, H_{\omega}]=  [\dX, V_{\omega}] \not=
  0$ for a.e.\ $\omega$, we have  $\cL \dX \not=0$, so   $\Chi_{\R\setminus \{0\}}(\cL) \dX\not= 0$. It follows that the measure $\Upsilon$ 
is nontrivial: 
\beq
\Upsilon(\R)=  \Upsilon(\R\setminus \{0\})= \tnorm[big]{\Chi_{\R\setminus \{0\}}(\cL) \dX}_2^{2} >0.
\eeq

\begin{lemma}\label{thmGamma>0}
Let $T>0$ and  ${\mu}\in\RR$. Then  
\beq\label{preGammaeq}
 \tfrac 1{4 T}  C_\mu^T \Chi_{\RR\setminus \set{0}}(\cL)    \le  - \cL_\perp^{-1}\cF_\mu^T \le   \tfrac 1{4T}  \Chi_{\RR\setminus \set{0}}(\cL),
\eeq
where 
\beq\label{CmuT}
 C_{{\mu}}^T := \inf_{E\in [E_{-}-E_+,E_{+}-E_-]}  \sech^{2}\pa{\tfrac {E-\mu}{T}}  >0,
\eeq
and $E_{\pm}$ are as in  \eq{Epm}. It follows that
\beq\label{Gammaeq}
\tfrac \pi {4T} C_{{\mu}}^T  \Upsilon(B) \le 
\Gamma_{{\mu}}^{T}(B) \le \tfrac \pi {4T} \Upsilon(B)  \qtx{for all Borels sets} B \subset \R,
\eeq
and 
the measure $\Gamma_{{\mu}}^{T}$ is nontrivial:   $\Gamma_{{\mu}}^{T}(\RR)>0$.

As a consequence, we have
\beq\label{slimT}
\slim_{T \to \infty} \Gamma_{{\mu}}^{T} = 0 \qtx{for all} {\mu}\in\RR. 
\eeq
\end{lemma}

\begin{proof}  Since  \eq{Gammaeq} follows from \eq{preGammaeq} using \cite[Eqs.~(2.43)]{KMu} and \eq{defUpsilon}, it suffices to prove \eq{preGammaeq}.

It follows from \cite[Eqs.~(2.39)--(2.40)]{KMu} that for $T>0$ we have
\begin{align}\label{supder}
- \cL_\perp^{-1}\cF_\mu^T& =F_\mu^T(\cH_L,\cH_R) \le \pa{\sup_{(\lambda_1,\lambda_2)\in \RR^2}F_\mu^T(\lambda_1,\lambda_2)} \Chi_{\RR\setminus \set{0}}(\cL) \\ 
\notag & \le \pa{\sup_{E\in \RR} \pa{-f_\mu^T(E)} ^\pr} \Chi_{\RR\setminus \set{0}}(\cL)= \tfrac 1 {4T}  \Chi_{\RR\setminus \set{0}}(\cL),
\end{align}
where we used the mean-value theorem.

The lower bound  is proved in a similar way.  We have 
\begin{align}
- \cL_\perp^{-1}\cF_\mu^T & =F_\mu^T(\cH_L,\cH_R)\ge \pa{\inf_{(\lambda_1,\lambda_2)\in [E_{-},E_{+}]^2}F_\mu^T(\lambda_1,\lambda_2) } \Chi_{\RR\setminus \set{0}}(\cL) \\ \notag & \ge  \pa{\inf_{E\in [E_{-}-E_+,E_{+}-E_-]} \pa{-f_\mu^T(E)} ^\pr }\Chi_{\RR\setminus \set{0}}(\cL)\ge    {\wtilde{C}_\mu^T } \Chi_{\RR\setminus \set{0}}(\cL),
\end{align}
where 
\begin{align}
\wtilde{C}_\mu^T  & := \tfrac 1 T \inf_{E\in [E_{-}-E_+,E_{+}-E_-]} \frac {\e^{\frac {E-\mu}{T}}}{\big(
\e^{\frac {E-\mu}{T}}+1 \big)^2}= \tfrac 1{4 T} C_\mu^T,
\end{align}
 with $C_\mu^T$ as in \eq{CmuT}.
\end{proof}

\begin{proof}[Proof of Theorem~\ref{thmTinfty}] 

It follows from \cite[Eqs.~(2.41)]{{KMu}} and \eq{supder} that
\beq
 \Psi\pa{\pa{-f_{{\mu}}^{T}}^{\pr}}\le \pi \tnorm[big]{\Chi_{ \{0\}}(\cL) \dX}_2^2  \sup_{E\in \RR} \pa{-f_\mu^T(E)} ^\pr \le \tfrac \pi {4T}\tnorm[big]{\Chi_{ \{0\}}(\cL) \dX}_2^2 ,
\eeq
so 
\beq \label{limPsi}
\lim_{T\to \infty} \Psi\pa{\pa{-f_{{\mu}}^{T}}^{\pr}}=0
 \qtx{for all} {\mu}\in\RR. 
 \eeq 
 
The theorem follows from \eq{decompose},   Lemma~ \ref{thmGamma>0}, and \eq{limPsi}.
\end{proof}

\section{Asymptotics with respect to the disorder}\label{secdisorder}

We now introduce a disorder parameter $\lambda\ge 0$. 
We consider the Anderson model $H_{\lambda}$, given by $H_{\omega, \lambda} := -\Delta + \lambda V_\omega$ (see \eq{AND}), and attach the label $\lambda$ to all quantities considered, when appropriate. (Note that $H_0= -\Delta$.)  We consider the small and large disorder limits of the conductivity measure.  Results of a similar nature in the BPK setting are given in \cite[Theorem~4.6]{BPK4}.
		
\begin{theorem}[Small disorder]
	\label{thm:small-dis}
	For  all $T >0$ and  $\mu\in\RR$ we have 
	\begin{equation}\label{limlambda0}
			\wwlim_{\lambda\to 0}   \Sigma_{\mu,\lambda}^{T} = \Sigma_{\mu,0}^{T}(\{0\})\, \boldsymbol{\delta}_{0} ,
	\end{equation}
	where 
	\beq\label{Sigma0mu}
	\Sigma_{\mu,0}^{T}(\{0\}) = \int_{\RR}\d\zeta \, (-f_{\mu}^{T})'(\zeta) 
	\Sigma_{\zeta,0}^{0}(\{0\}) =  \int_{\RR}\d\zeta \, (-f_{\mu}^{T})'(\zeta) 
	\psi(\zeta)>0.
	\eeq
\end{theorem}

\begin{proof}
 To  prove \eq{limlambda0}  it suffices  	to show convergence of the Fourier transforms for every $t\in\RR$. Given  $T>0$ and  $\mu\in\RR$,  we have
	\begin{align}
 		\int_{\RR}\! \Sigma_{\mu,\lambda}^{T}(\d\nu) \, \e^{{i}t \nu} &= 
		\pi \big\langle\mkern-4mu\big\langle \e^{{i}t\cL_{\lambda}} \dot{X}_{1}, Y_{\mu}^{T}\big\rangle\mkern-4mu\big\rangle
		 = i\pi \big\langle\mkern-4mu\big\langle \e^{{i}t\cL_{\lambda}} \dot{X}_{1}, [X_{1}, f_{\mu}^{T}(H_{\lambda})]\big\rangle\mkern-4mu\big\rangle \nonumber\\
		 &= {i}\EE\Big\{ \big\langle \e^{{i}t H_{\omega,\lambda}} \dot{X}_{1} \e^{-{i}t H_{\omega,\lambda}} \delta_{0}, 
		 X_{1} f_{\mu}^{T}(H_{\omega,\lambda}) \delta_{0} \big\rangle\Big\} .
	\end{align}
	Since in view of \eq{vpm} we have  
	\beq
	\norm{H_{\omega,\lambda} - (-\Delta)}=\norm{\lambda V_\omega} \le \lambda \max \set{\abs{v_-}, \abs{v_+}} \quad \text{with probability one},
	\eeq  
		we get
\begin{align}
	\lim_{\lambda\to 0 } \int_{\RR}\! \Sigma_{\mu,\lambda}^{T}(\d\nu) \, \e^{{i}t \nu}
	&=  {i}\EE\Big\{ \big\langle \e^{{i}t (-\Delta)} \dot{X}_{1} \e^{-{i}t (-\Delta)} \delta_{0}, 
		 X_{1} f_{\mu}^{T}(-\Delta) \delta_{0} \big\rangle\Big\} \nonumber\\
	&= 	 \int_{\RR}\! \Sigma_{\mu,0}^{T}(\d\nu) \, \e^{{i}t \nu} = \Sigma_{\mu,0}^{T}(\{0\}),
 \end{align}
where we used $\Sigma_{\mu,0}^{T} = \Sigma_{\mu,0}^{T}(\{0\}) \,\boldsymbol{\delta}_{0}$, which follows  from \eq{convolution}   and  \cite[Eq.\ (2.53)]{KMu}, which also imply \eq{Sigma0mu} .
\end{proof}

\begin{theorem}[Large disorder]
\label{thm:large} Given  $T\ge 0$ and  $\mu\in\RR$, there exist $\lambda_2<\infty $ and a finite constant $C$ independent of $\lambda$,  such that
\beq\label{SigmaRbound0}
 \Sigma_{\mu,\lambda}^{T} (\RR) \le  C \lambda^{-\frac 14}  \qtx{for all} \lambda \ge \lambda_2.
\eeq
In particular, we have
		\begin{equation}\label{lambdatoinfty}
			\slim_{\lambda\to \infty } \Sigma_{\mu,\lambda}^{T} = 0  .
				\end{equation}	
\end{theorem}
	
 \begin{proof}
  Fix $T\ge 0$ and  $\mu \in \R$.
	Recall that there exists $\lambda_0 >0$ such that  $ \Xi_0=\RR$ for $\lambda \ge \lambda_0$.  Thus  it follows from \eq{kubo-expr}, proceeding as in  \cite[Eq. (3.19)]{KMu}, that   for all $\lambda \ge \lambda_0$ we have ($\widehat{x}_{1}$ denotes  the unit vector in the
$x_{1}$-direction) \begin{align}\label{Sigma31999}
 \Sigma_{\mu,\lambda}^{T} (\RR)  & =  \pi \llangle \dot{X}_{1},
    Y_{\mu,\lambda}^{T} \rrangle= - \pi\mathbb{E} \bigl\{ \langle X_ 
  {1}^{2} H_{\omega,\lambda}
  \delta_{0}, f_{\mu}^{T}(H_{\omega,\lambda}
)\delta_{0}\rangle \bigr\} \\
&=  - \pi  \mathbb{E} \bigl\{ \langle \delta_{\widehat{x}_{1}} +
  \delta_{-\widehat{x}_{1}}, f_{\mu}^{T}(H_{\omega,\lambda})\delta_{0}\rangle \bigr 
  \}= -2\pi \Re \mathbb{E} \bigl\{ \langle \delta_{\widehat{x}_{1}} , f_{\mu}^{T}(H_{\omega,\lambda})\delta_{0}\rangle \bigr  \}\notag,
  \end{align}
where we used covariance for the last equality.  In particular, we have
\beq\label{SigmaRbound}
 \Sigma_{\mu,\lambda}^{T} (\RR) \le 2\pi\abs{ \mathbb{E} \bigl\{ \langle \delta_{\widehat{x}_{1}} , f_{\mu}^{T}(H_{\omega,\lambda})\delta_{0}\rangle \bigr  \}}.
\eeq

 Let  $\wtilde{H}_{\omega,\lambda} 
		:= - \lambda^{-1} \Delta + V_{\omega}$, so $H_{\omega,\lambda} =\lambda \wtilde{H}_{\omega,\lambda}$.  Note that   $\wtilde{\mathfrak{S}}_\lambda := \frac 1 \lambda\mathfrak{S}_\lambda = \sigma (\wtilde{H}_{\omega,\lambda} )$ with probability one. 
	We have 
	\beq
	\label{F-l-def}
	f_{\mu}^{T}(H_{\omega,\lambda})  
		=F_{\lambda}(\wtilde{H}_{\omega,\lambda}), \qtx{where} F_{\lambda} := f_{\frac {\mu} {\lambda}}^{\frac T \lambda}.
		\eeq

Without loss of generality we take $\lambda\ge \lambda_0$ large enough  to ensure (recall\eq{vpm}-\eq{Epm})  
\beq\label{mulambda}
\wtilde{\mathfrak{S}}_\lambda \subset [v_- -1, v_+ +1]  \qtx{and} {\abs{\mu}}\le \lambda^{\frac 34}
 	\quad(\text{i.e.,}  \tfrac {\abs{\mu}} {\lambda}   \le	 \lambda^{-\frac 14}).
	\eeq
	We fix a ($\lambda$-independent) function  $h \in C_{c}^\infty(\RR)$, $0\le h\le 1$, such that 
	\begin{align}\label{derivh-bound}
& h(s) := 1 \mqtx{if} \;   s \in  [v_- -1, v_+ +1], \quad h(s)  := 0  \mqtx{if}  s \notin    [v_- -3, v_+ +3], 
\notag\\
&\big| h^{(r)}\big|   \le  \gamma \Chi_{ [v_- -3, v_+ +3]\setminus  [v_- -1, v_+ +1]}  \qtx{for}  r=1,2,3
\end{align}
where  $\gamma>0$ is some universal constant.
Note that 
 \beq\label{defJ}
 J_{\lambda}(\wtilde{H}_{\omega,\lambda})= F_{\lambda}(\wtilde{H}_{\omega,\lambda}), \qtx{where} J_\lambda := hF_\lambda.
 \eeq

 {For each $\lambda$ we  fix
an even function $g_{\lambda} \in C_{c}^\infty(\RR)$, $0\le g_\lambda\le 1$, such that
\begin{align}\label{deriv-bound}
&g_\lambda(s) := 1 \; \text{\quad if  }  \;   \abs{s} \le    3  \lambda^{-\frac 14}, \quad g_\lambda(s) := 0  \; \text{\quad if }  \;   \abs{s} \ge   5   \lambda^{-\frac 14},   \notag\\
&\big|g_\lambda^{(r)}\big|   \le  \gamma \lambda^{\frac {r}4}  \Chi_{[- 5   \lambda^{-\frac 14}, 5   \lambda^{-\frac 14}]} \qtx{for} r=1,2,3.
\end{align}
We write
 \beq
J_{\lambda} =  J_{1,\lambda}+  J_{2,\lambda}, \qtx{where} J_{1,\lambda} := g_\lambda J_\lambda  \qtx{and} J_{2,\lambda} :=  (1 -g_{\lambda}) J_{\lambda} .
 \eeq
 }
  
  Let $\wtilde{\mathcal{N}}_\lambda$ denote the density of states measure for $\wtilde H_{\omega,\lambda}$, i.e.,
\beq
\wtilde{\mathcal{N}}_\lambda (B) := \EE\set{ \langle \delta_0,\Chi_B (\wtilde H_{\omega,\lambda} )\delta_0 \rangle} \qtx{for Borel sets} B\subset \R.
\eeq  
It follows from the Wegner estimate  that $\wtilde{\mathcal{N}}_\lambda$ is absolutely continuous, and its Lebesgue density obeys the $\lambda$-independent bound $\frac {\d \wtilde{\mathcal{N}}_\lambda}{\d E} \le \norm{\rho}_\infty$.  
{Thus, we get, using also  $0\le J_{1,\lambda} \le \Chi_{[- 5   \lambda^{-\frac 14},\ 5   \lambda^{-\frac 14}]}$ (recall $0\le F_\lambda \le 1$), 
\begin{align}\notag
&\Big|\EE\set{\langle \delta_{\widehat{x}_{1}},{J_{1,\lambda}(\wtilde H_{\omega,\lambda} )}\delta_0 \rangle}\Big| \le \EE \set{\Big\| \pa{J_{1,\lambda}(\wtilde H_{\omega,\lambda} )}^{\frac 12} \delta_{\widehat{x}_{1}}\Big\| 
	\Big\|\pa{J_{1,\lambda}(\wtilde H_{\omega,\lambda} )}^{\frac 12} \delta_{0}\Big\| }
 \\ & \qquad \qquad \le  \EE \set{ \Big\|\pa{J_{1,\lambda}(\wtilde H_{\omega,\lambda} )}^{\frac 12} \delta_{0} \Big\|^2 }
= \EE\set{ \langle \delta_0,{J_{1,\lambda}(\wtilde H_{\omega,\lambda} )}\delta_0 \rangle}  \label{useWeg2} \notag\\
& \qquad \qquad = \int_{\RR}\d \wtilde{\mathcal{N}}_\lambda(E) {J_{1,\lambda}(E )} \le \norm{\rho}_\infty
		\int_{\RR} \d E \,  {J_{1,\lambda}(E )}\le  {10 \norm{\rho}_\infty}   \lambda^{-\frac 14}. 
\end{align}}

We now estimate $\EE\set{\langle \delta_{\widehat{x}_{1}},{J_{2,\lambda}(\wtilde H_{\omega,\lambda} )}\delta_0 \rangle}$. Note that for  any bounded measurable function $k$ on $\R$ we have
\beq
\langle \delta_{\widehat{x}_{1}} , k(V_{\omega})\delta_{0}\rangle=  \langle \delta_{\widehat{x}_{1}} , \delta_{0}\rangle  =  0.
\eeq
Thus, with probability one we have
\begin{align}
 	\langle \delta_{\widehat{x}_{1}},{J_{2,\lambda}(\wtilde H_{\omega,\lambda} )}\delta_0 \rangle 
	&= \langle \delta_{\widehat{x}_{1}},\big( {J_{2,\lambda}(\wtilde H_{\omega,\lambda} )} - J_{2,\lambda}(V_{\omega}) \big)\delta_0 \rangle \notag\\
	&= \frac{1}{\lambda} \int_{\CC} \d\wtilde{J_{2,\lambda}}(z) \, \langle\delta_{\widehat{x}_{1}},\frac{1}{\wtilde{H}_{\omega,\lambda} -z} \,\Delta\,
	\frac{1}{V_{\omega} -z} \delta_0 \rangle ,
\end{align}
where the second equality relies on the Helffer-Sj\"ostrand formula. (We refer to \cite[App.\ B]{HuSi00} for a review of  the Helffer-Sj\"ostrand formula.) Given a smooth function $\zeta$  on the real line,   $\wtilde{\zeta}$ denotes an almost analytic extension of $\zeta$ to the complex plane.  We recall the estimate
\begin{equation}
  \int_{\CC} \! |\d\wtilde{\zeta}(z)| \;\frac{1}{|\mathrm{Im}\, z|^2}  \le c_3 \;  \{\!\{\zeta\}\!\}_3 ,
\end{equation}
where $c_3$ is a finite constant independent of the function $\zeta$, and 
\begin{equation} 
	\label{HSnorm}
  \{\!\{\zeta\}\!\}_3 := \sum_{r=0}^3 \int_{\mathbb{R}}\!\mathrm{d}s\;
  |\zeta^{(r)}(s)|\,(1 + \abs{s}^{2})^{\frac {r-1} 2} . 
\end{equation}
We thus conclude that
\begin{align}\label{estbyHS}
 	\big|\langle \delta_{\widehat{x}_{1}},{J_{2,\lambda}(\wtilde H_{\omega,\lambda} )}\delta_0 \rangle \big| 
	\le  \frac{2d c_3}{\lambda} \{\!\{J_{2,\lambda}\}\!\}_3,
\end{align}
and need to 
 estimate $\{\!\{J_{2,\lambda}\}\!\}_3$. If $T=0$ we have $F^{\prime}_\lambda (E)=0$ for
$\abs{E} \ge 3   \lambda^{-\frac 14}$.
If  $T>0$ we have 
 \begin{align}
 \abs{F^\pr_\lambda (E)}&=\tfrac {\lambda} {2T} \sech^2 \pa{\tfrac {\lambda} {2T} (E-\tfrac \mu \lambda)}\le \tfrac {\lambda} {2T} \sech^2 \pa{\tfrac {\lambda} {2T} (2   \lambda^{-\frac 14})}\\ \notag
 & =
  \tfrac {\lambda} {2T} \sech^2 \pa{\tfrac { \lambda^{\frac 34}}  {T}} \qtx{for} \abs{E} \ge  3   \lambda^{-\frac 14},
  \end{align}
 with similar estimates for $|F^{(r)}_\lambda (E)|$, $r=2,3$. (We also used \eq{mulambda}.)  We conclude that
there is a constant $K>0$ such that
 \beq
 \max_{r=1,2,3} |F^{(r)}_\lambda (E)| \le K < \infty \qtx{for} \abs{E} \ge 3    \lambda^{-\frac 14}.
 \eeq
 Combining with  \eq{mulambda}, \eq{derivh-bound}, and  \eq{deriv-bound} we conclude that
 \beq\label{HS3J}
 \{\!\{J_{2,\lambda}\}\!\}_3 \le K'  \lambda^{\frac 34}, \sqtx{where} K' \;\text{is a finite constant independent of} \; \lambda.
 \eeq

{Combining  \eq{estbyHS} and \eq{HS3J}  we get the deterministic bound
\begin{align}\label{usedecay}
 	\abs{\langle \delta_{\widehat{x}_{1}},{J_{2,\lambda}(\wtilde H_{\omega,\lambda} )}\delta_0 \rangle} \le  2d c_3 K'   \lambda^{-\frac 14} ,
\end{align}
valid on an event of probability one.}

Taking the expectation of \eqref{usedecay}, and using \eq{SigmaRbound},  \eq {defJ} and  \eqref{useWeg2}, we get 
\beq\label{SigmaRbound2}
 \Sigma_{\mu,\lambda}^{T} (\RR) \le  C \lambda^{-\frac 14}  \qtx{for all} \lambda \ge \lambda_2,
\eeq
where $\lambda_2<\infty $ and $C$ is finite constant independent of $\lambda$.
\end{proof}

\section{Electromagnetic energy absorption}\label{secwork}

We now consider the Anderson model as in \eq{AND}  and  an electric field as in \eq{assumption1}    such that ${\cE} \in \mathrm{L}^1(\RR)$.  Notice that this is equivalent to  assume
\beq\label{assumption2}
\cE(t) =  \int_{\RR}\!\d \nu \; \e^{i\nu t}\widehat{\cE}(\nu), \mqtx{where} \cE,  \widehat{\cE} \in  \mathrm{L}^1(\RR)  \mqtx{with}
   \widehat{\cE}(\nu)=\overline{ \widehat{\cE}(-\nu)}.
   \eeq
Note that 	\eq{assumption2} implies
$\cE,  \widehat{\cE} \in  C(\RR) \cap \mathrm{L}^1(\RR)\cap \mathrm{L}^\infty(\RR)$.

 With the extra assumption of ${\cE} \in \mathrm{L}^1(\RR)$  we can proceed without adiabatic switching, i.e., with $\eta=0$.
  If  $T>0$ or  ${\mu}\in \Xi_0$,   the energy of the system at time $t$ is given by $\cT (H_{\omega} (t) \varrho_{{\mu},\omega}^T (t) )$.  The total energy the solid absorbs during all times from the electric field is given by 
\begin{align}\notag
 	W_{{\mu}}^T(\cE) &=  \lim_{t\to\infty}  \cT \big(H_{\omega} (t) \varrho_{{\mu},\omega}^T (t) \big)  - \lim_{t\to-\infty}  \cT \big(H_{\omega} (t) \varrho_{{\mu},\omega}^T (t) \big) \\
	&= \int_{\RR}\!\d t \, \tfrac{\d}{\d t} \,\cT \big(H_{\omega} (t) \varrho_{{\mu},\omega}^T (t) \big)\\&= \int_{\RR}\!\d t \, \Big( \cT\big(  - i[H_{\omega} (t), \varrho_{{\mu},\omega}^T (t) ] H_{\omega} (t)\big)  + 
			 		\cT\big(  \varrho_{{\mu},\omega}^T (t) \tfrac{\d}{\d t} H_{\omega} (t) \big)   \Big) \nonumber\\
		&= {i}\int_{\RR}\!\d t \,  \cE(t) \cT\big( \varrho_{{\mu},\omega}^T (t)[X_{1}, H_{\omega} (t)]\big)  
		 =  \int_{\RR}\!\d t \,  \cE(t) J(t,{\mu},T,\cE).
\end{align}
Here we used \eq{Liouvilleeq},   the cyclic invariance of $\cT$ to get
\beq
\cT\big(  - i[H_{\omega} (t), \varrho_{{\mu},\omega}^T (t) ] H_{\omega} (t)\big)  =0,
\eeq
as well as \eq{Homegaeta}, \eq{dotX} and \eqref{curdef}.
Note that  $J(t,\mu,T,\cE)$, and hence $W_{{\mu}}^T(\cE), $ are  real-valued since $H_\omega(t)$ is self-adjoint  and  $ \varrho_{{\mu},\omega}^T (t)\ge 0$.

The absorption of electromagnetic energy in linear response theory can now be seen to be well defined in terms of the linear-response current $J_{\mathrm{lin}}(t;{{\mu}},T,\cE)$ (see \eq{lin-cur}  and \eq{lin-cur22}):
\begin{align}\label{linearheat}
 	W^T_{\mathrm{{\mu},lin}}(\cE) :\!&= \lim_{\alpha\to 0} \, \frac{W_{{\mu}}^T(\alpha \cE)}{\alpha^{2}}
		=  \int_{\RR}\!\d t \,  \cE(t)  J_{\mathrm{lin}}(t;{{\mu}},T,\cE)\\ 
		\notag
		& = \frac{1}{\pi}  \int_{\RR}\!\d t \,  \cE(t)    \int_{-\infty}^{t}\!\d s\;  \cE(s)  \int_{\mathbb{R}}\!\Sigma_{{\mu}}^{T}(\d\nu)\  \e^{- i (t-s) \nu} \\
		\notag
		& =\frac{1}{\pi} \int_{\RR^2}\!\d t \d s \,  \cE(t)  \cE(s) \Chi_{[0, \infty[}(t-s) \int_{\mathbb{R}}\!\Sigma_{{\mu}}^{T}(\d\nu)\e^{- i (t-s) \nu}\\
    \notag
		& =\frac{1}{\pi} \int_{\RR^2}\!\d t \d s \,  \cE(t)  \cE(s) \Chi_{[0, \infty[}(t-s) \int_{\mathbb{R}}\!\Sigma_{{\mu}}^{T}(\d\nu)\cos \big((t-s) \nu\big)  \\
    \notag
		& =\frac{1}{2\pi} \int_{\RR^2}\!\d t \d s \,  \cE(t)  \cE(s)\int_{\mathbb{R}}\!\Sigma_{{\mu}}^{T}(\d\nu)\cos \big((t-s) \nu\big)  \\
    \notag
		& =\frac{1}{2\pi} \int_{\RR^2}\!\d t \d s \,  \cE(t)  \cE(s)\int_{\mathbb{R}}\!\Sigma_{{\mu}}^{T}(\d\nu)\e^{- i (t-s) \nu}\\
		\notag
		&=  2\pi \int_{\RR}\!\Sigma_{{\mu}}^{T}(\d\nu)  \,  |\what{\cE}(\nu)|^{2},
\end{align}
where we used the fact that $\Sigma_{{\mu}}^{T}$ is an even measure.

Thus we proved the following theorem. (See \cite[Theorem~4.7]{BPK4} for an analogous result in the BPK setting.)

  \begin{theorem} [Electromagnetic energy absorption]  \label{thm:heat}  Consider the Anderson model as in \eq{AND}  and  an electric field \bu{$\cE$} as in \eq{assumption2}.   Suppose  either $T>0$ or  ${\mu}\in \Xi_0$.  Then 	
\begin{align}
 		W^T_{\mathrm{{\mu},lin}}(\cE)
		= 2\pi\int_{\RR}\!\Sigma_{{\mu}}^{T}(\d\nu)  \, |\what{\cE}(\nu)|^{2} \ge 0 . \label{heat>}
	\end{align}  
In particular, we have	
\begin{align}
 		W^T_{\mathrm{{\mu},lin}}(\cE)
		= 2\pi\int_{\RR}\!\Gamma_{{\mu}}^{T}(\d\nu)  \, |\what{\cE}(\nu)|^{2} \ge 0  \qtx{if} \what{\cE}(0)= 2\pi \int_{\R}\d t\, \cE(t)=0. \label{heat>0}
	\end{align}  
		
	In addition, for all $T>0$ and  $\mu \in\R$ we have
	\beq\label{W>0}
	W^T_{\mathrm{{\mu},lin}}(\cE)\ge  2\pi\int_{\RR}\!\Gamma_{{\mu}}^{T}(\d\nu)  \, |\what{\cE}(\nu)|^{2}\ge \tfrac {\pi^2} {2T}  C_{{\mu}}^T\int_{\RR}\!\Upsilon(\d\nu)  \, |\what{\cE}(\nu)|^{2}  .
	\eeq	
\end{theorem}

\begin{remark}  It follows from  Remark~\ref{SigmasigmacL} that
only  frequencies $\nu\in \sigma (\cL)$ contribute to  $W^T_{\mathrm{{\mu},lin}}(\cE)$. In particular, we have
\beq
W^T_{\mathrm{{\mu},lin}}(\cE)
		= 2\pi\int_{\sigma(\cL)}\!\Sigma_{{\mu}}^{T}(\d\nu)  \, |\what{\cE}(\nu)|^{2} \qtx{if either} T>0 \qtx{or}{\mu}\in \Xi_0.
\eeq
\end{remark}

\end{document}